\documentclass{article}
\usepackage{spconf,amsmath,graphicx}
\usepackage{epstopdf}
\usepackage{multirow, makecell}
\usepackage{cite}
\usepackage{lscape}

\makeatletter
\def\set@curr@file#1{%
  \begingroup
    \escapechar\m@ne
    \xdef\@curr@file{\expandafter\string\csname #1\endcsname}%
  \endgroup
}
\def\quote@name#1{"\quote@@name#1\@gobble""}
\def\quote@@name#1"{#1\quote@@name}
\def\unquote@name#1{\quote@@name#1\@gobble"}
\makeatother

\title{Classification of High-Dimensional Motor Imagery Tasks based on An End-to-end role assigned convolutional neural network}
\name{Byeong-Hoo Lee$^1$ \qquad  Ji-Hoon Jeong$^1$ \qquad Kyung-Hwan Shim$^1$ \qquad Seong-Whan Lee$^{1,2}$\thanks{This work was partly supported by Institute of Information \& Communications Technology Planning \& Evaluation (IITP) grant funded by the Korea government (No. 2017-0-00432, Development of Non-Invasive Integrated BCI SW Platform to Control Home Appliances and External Devices by User’s Thought via AR/VR Interface) and partly funded by Institute of Information \& Communications Technology Planning \& Evaluation (IITP) grant funded by the Korea government (No. 2017-0-00451, Development of BCI based Brain and Cognitive Computing Technology for Recognizing User’s Intentions using Deep Learning).}}
\address{$^1$Department of Brain and Cognitive Engineering, Korea University\\$^2$Department of Artificial Intelligence, Korea University}
\begin{document}
%
\maketitle
\begin{abstract}
A brain-computer interface (BCI) provides a direct communication pathway between user and external devices. Electroencephalogram (EEG) motor imagery (MI) paradigm is widely used in non-invasive BCI to obtain encoded signals contained user intention of movement execution. However, EEG has intricate and non-stationary properties resulting in insufficient decoding performance. By imagining numerous movements of a single-arm, decoding performance can be improved without artificial command matching. In this study, we collected intuitive EEG data contained the nine different types of movements of a single-arm from 9 subjects. We propose an end-to-end role assigned convolutional neural network (ERA-CNN) which considers discriminative features of each upper limb region by adopting the principle of a hierarchical CNN architecture. The proposed model outperforms previous methods on 3-class, 5-class and two different types of 7-class classification tasks. Hence, we demonstrate the possibility of decoding user intention by using only EEG signals with robust performance using an ERA-CNN.

\end{abstract}
\begin{keywords}
Brain-computer interface (BCI), Electroencephalogram (EEG), Motor imagery, Convolutional Neural Network (CNN)
\end{keywords}

\section{Introduction}
\label{sec:intro}
Brain-computer interface (BCI) has been studied for motor-disabled patients to recover and replace their motor function, and even for healthy users to extend motor function capabilities with external devices control \cite{C1,C2,C3}. In non-invasive BCI paradigms, EEG signals are easily collected without brain surgery and commonly used due to their high temporal resolution \cite{C4}. The EEG signals have been applied to various types of BCI paradigms such as event-related potential (ERP) \cite{ERP}, movement-related cortical potential (MRCP) \cite{MRCP} and motor imagery (MI) \cite{MI}. EEG-based BCI paradigms have been developed for interaction between users and external devices \cite{speller, roboticarm, drone, wheelchair}. Of these paradigms, MI-based BCI decodes the EEG signals when the user imagines movements. While the user performs a MI task, event-related desynchronization/synchronization (ERD/ERS) patterns represented spectral features over the supplementary motor area and pre-motor cortex \cite{ERD}.

Decoding user intention from EEG data is one of the most challenging issues of BCI. One of the main reasons is that EEG signals have intricate and non-stationary properties and low signal quality \cite{nonstationary}. For MI, it is especially difficult to obtain high-quality data, as it is unknown what the user exactly imagined. Therefore, recent advances related to MI-based BCI approaches have investigated for improving the decoding accuracy using numerous feature extraction or classification methods based on advanced machine learning algorithms and deep learning. For example, filter bank common spatial pattern (FBCSP) algorithm \cite{FBCSP} has been widely adopted for MI classification with linear discriminant analysis (LDA) using spectral power modulations \cite{channel,SSVEP2}. Inspired by FBCSP, deep and shallow convolutional neural network (CNN) was developed for finding causal contributions of features in the different frequency bands \cite{deepconvnet}. A compact CNN with depthwise convolution is trained to summarize individual feature maps over time to classify EEG data \cite{EEGNET}. These studies focused primarily on a few classes classification and non-intuitive tasks contained in BCI Competition IV data (left-hand, right-hand, foot, and tongue). However, intuitive MI is a practical BCI paradigm due to direct interaction between users and devices without artificial command matching \cite{intuitiveMI}. To the best of our knowledge, these approaches have not achieved satisfactory classification performance on intuitive MI yet. Therefore, in this paper, we focused on intuitive MI data classification containing various types of movements of a single-arm.

Hence, our main contributions represented in three folds: 1) We collected EEG data concerning single-arm movement imagery; arm reaching task in 3D space, hand grasping, and wrist-twisting. 2) We proposed an end-to-end role-assigned CNN (ERA-CNN) for classifying various MI tasks with high performance by adopting the principle of a hierarchical CNN architecture which extracts discriminative features from different body regions such as the arm, hand, and wrist of a single-arm. 3) The proposed ERA-CNN model achieved substantial improvement in MI classification performance, and we have shown that the principle of hierarchy is efficient at uncertain multi-class data classification.

\begin{figure}[!t]
  \centerline{\includegraphics[width = \columnwidth]{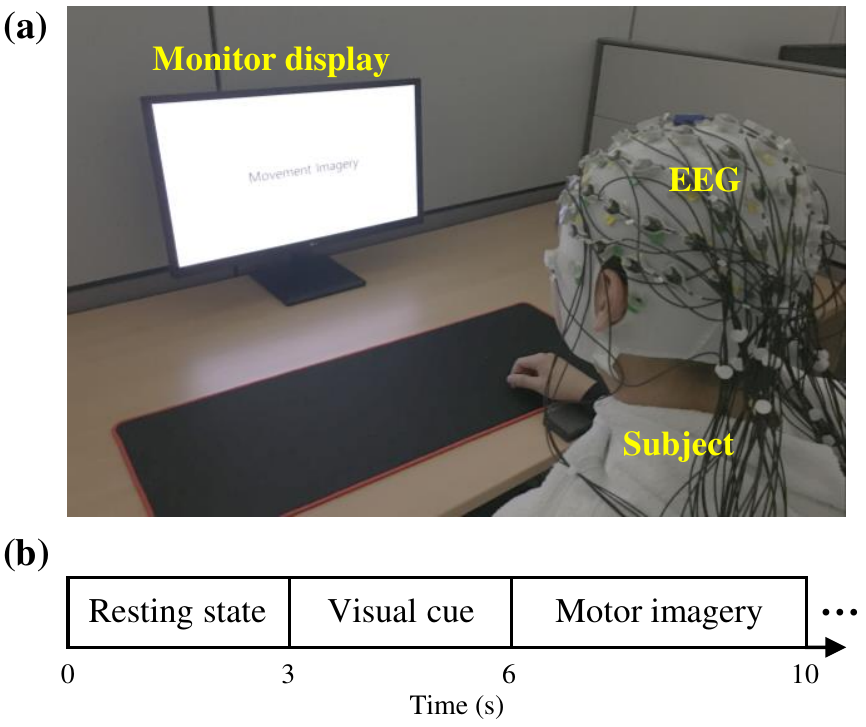}}
  \caption{(a) Experimental environment for EEG data acquisition, (b) Experimental paradigm for a single-trial}
 \label{fig_paradigm}
\end{figure}

\section{Methods}
\label{sec:format}

\subsection{Data description}
\label{sec:pagestyle}
We collected an intuitive MI dataset from nine healthy subjects between the age of 22 and 30 (6 males and 3 females, all right-handed). An EEG signal amplifier (BrainAmp, BrainProduct GmbH, Germany) was selected to recordEEG signals with a sampling rate of 1000 Hz and a 60 Hz notch filter. Additionally, a band-pass filter from 1-60 Hz applied to all channels. BrainVision software was used for data recording with 64 Ag/AgCl electrodes according to 10-20 international system. The FPz and FCz channels were selected as ground and reference respectively. From these 64 channels, we selected 24 channels placed on the motor cortex \cite{channel}, which are most relevant for the MI task (F3, F1, Fz, F2, F4, FC3, FC1, FC2, FC4, C3, C1, Cz, C2, C4, CP3, CP1, CPz, CP2, CP4, P3, P1, Pz, P2, and P4). Impedances were measured between the electrodes and the scalp to maintain channels impedance below 15 k$\Omega$. During the experiment, subjects were asked to imagine specific muscle movements following the paradigm in Fig. \ref{fig_paradigm} and performed 50 trials per each task. Total 9 classes of single-arm tasks were defined: arm-reaching (left, right, forward, backward, upward, downward), grasping, twisting, and the resting state. We divided the 6-class of arm-reaching tasks into horizontal reaching and vertical reaching. Additionally, the dataset was resampled at 250 Hz before classification. Data validation was done using an FBCSP and regularized linear discriminant analysis (RLDA) for each class. The protocols and environments were reviewed and approved by the Institutional Review Board at Korea University [1040548-KU-IRB-17-172-A-2].

\subsection{ERA-CNN}
\label{sec:typestyle}
ERA-CNN is an end-to-end convolutional neural network designed to extract frequency features through hierarchical convolution layers. Generally, hierarchical CNN consists of a shared layer and several sub-networks to separately obtain higher-level features \cite{hier}. In the following section, we describe the design choices and training strategy of ERA-CNN. The overview of our architecture is shown in Fig. 2.

\begin{figure*}[!t]
  \centerline{\includegraphics[scale = 0.82]{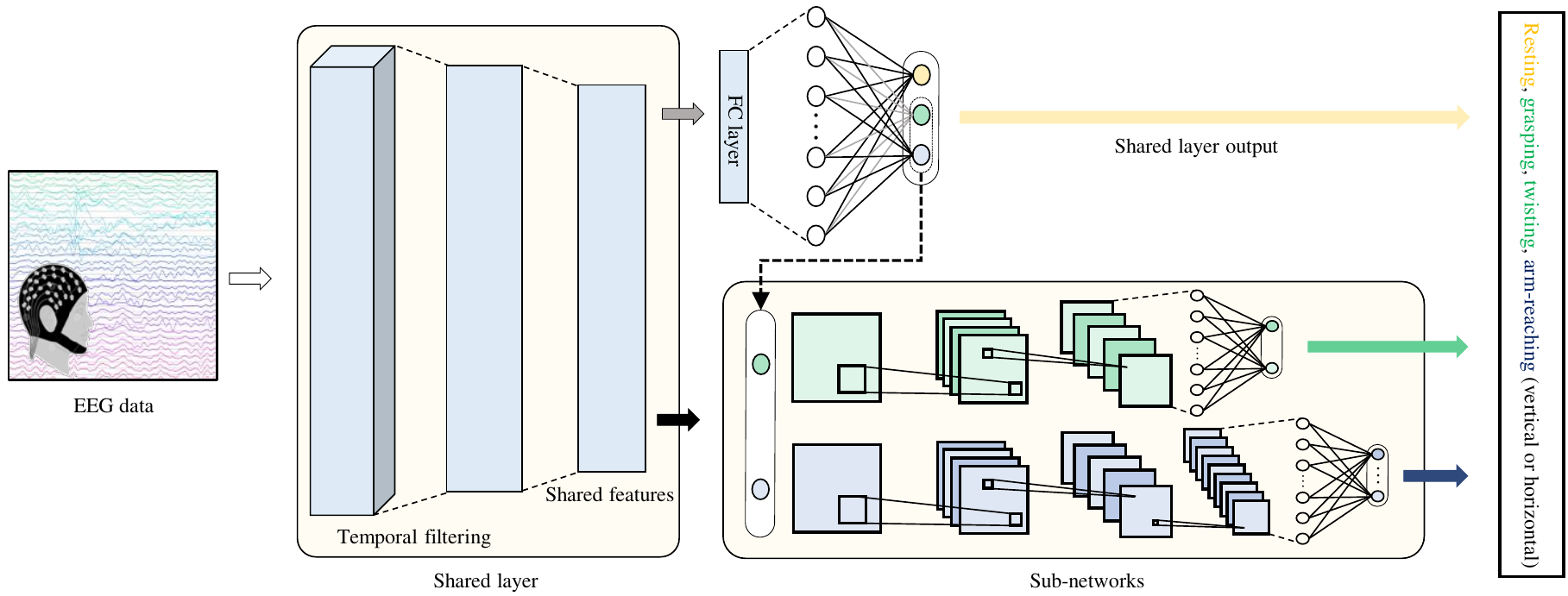}}
  \caption{Overall flowchart of the proposed ERA-CNN}
\end{figure*}

\subsubsection{Shared layer for raw EEG signals}
\label{sssec:subsubhead}

The shared layer consisted of two convolution blocks which classify each category. The first convolution block is composed of a temporal convolution layer and a spatial filter layer to reduce the dimensionality to a single channel. The temporal kernel size is set to a quarter of the input's sampling rate (which creates a receptive field above 4 Hz) to remove ocular artifacts. In the second block, the convolution layer and softmax function conduct the categorization of classes defining it as arm-reaching MI or hand-related MI (grasping and twisting) or the resting state. If prediction of shared layer is not resting state, sub-networks utilize the features (shared features) from the second convolution layer of the shared layer as input to conduct detailed classifications for each sub-category.

\subsubsection{Sub-networks}
\label{sssec:subsubhead}
Two sub-networks were exploited to improve classification accuracy, each specializing in predicting different types of MI tasks. A sub-network for hand-related MI classification is composed of three convolution-pooling blocks. Similarly, a sub-network for arm-reaching MI classification consists of four convolution-pooling blocks with a smaller kernel size to extract features. An extra block was added since more classes have to be classified using the same amount of shared features. The softmax function was applied to provide the final classification of each sub-network.

Contrary to other hierarchical CNNs, both sub-networks received shared features regardless of the shared layer output during training. In this way, one sub-network learns the correct classification, while the other sub-network learns the wrong cases at the same time. By training both cases, the sub-networks specialized in each classification role. In every convolution block of the ERA-CNN, we applied average pooling in order to reduce the dimensionality and perform smoothing of the EEG data. The exponential linear unit (ELU) was applied as the activation function\cite{ELU}, which can help avoid severe distortion of EEG data. The detailed design choices and filter sizes are described in Table 1.

\begin{table}[t!]
{\normalsize
\caption{Design choices of ERA-CNN}
\renewcommand{\arraystretch}{1.23}
\resizebox{\columnwidth}{!}{%
\begin{tabular}{llll} \Xhline{4\arrayrulewidth}
Parameter                     & Shared layer        & Sub-network (Arm)        & Sub-network (Hand)       \\ \hline
\multirow{3}{*}{Input}        & Raw EEG             & Shared features           & Shared features           \\ 
                              & (1, 1, 24, 751)     & (1, 36, 1, 216)          & (1, 36, 1, 216)          \\
\multirow{4}{*}{Hidden layer} & Conv2D: 36          & Conv2D: 36, 72, 144, 288 & Conv2D: 36, 72, 144, 288 \\ \hline
                              & AvgPool: (1,3)      & AvgPool: (1,3)           & AvgPool: (1,3)           \\
                              & Stride: (1,3)       & Stride: (1,3)            & Stride: (1,3)            \\ \hline
\multirow{2}{*}{Activation}   & ELU                 & ELU                      & ELU                      \\
                              & Last layer: Softmax & Last layer: Softmax      & Last layer: Softmax      \\ \hline
Optimizer                     & Adam                & Adam                     & Adam                     \\ \hline
Loss                          & Cross entropy       & Cross entropy            & Cross entropy      \\     \Xhline{4\arrayrulewidth}
\end{tabular}}}
\end{table}

\subsubsection{Loss functions}
\label{sssec:subsubhead}

The ERA-CNN loss function consisted of three separate terms and the output of a shared layer is the probabilities for each categorized class. ERA-CNN selects sub-networks based on the prediction probability of each class. In order to take into account the uncertainty in this prediction for each selection (i.e. contribution to sub-networks of the shared layer), the loss function was modified as follows:
\begin{equation}
loss(L_s, L_a, L_h) = p_a L_a + p_h L_h + L_s
\end{equation}
where ${p_a}$ is a probability to select a sub-network for arm-reaching MI classification and ${p_h}$ is a  probability to select a sub-network for hand-related MI classification. ${L_s}$, ${L_a}$ and ${L_h}$ are the loss of the shared layer, sub-network for arm-reaching MI classification and sub-network for hand-related MI classification respectively. These loss values are derived from the cross-entropy loss function \cite{entropy} which is a weighted sum of loss values as:
\begin{equation}
L_s = -\sum_{c=1}^{3}y_{s.c}\log{\hat{y}_{s.c}}
\end{equation}
\begin{equation}
L_a = -\sum_{c=1}^{M}y_{a.c}\log{\hat{y}_{a.c}}
\end{equation}
\begin{equation}
L_h = -\sum_{c=1}^{2}y_{h.c}\log{\hat{y}_{h.c}}
\end{equation}
where ${y_{s.c}}$ is label of the shared layer, and ${y_{a.c}}$ and ${y_{h.c}}$ are labels of the arm-reaching and hand-related MI respectively. ${\hat{y}_{s.c}}$ is the classification output of a shared layer. ${\hat{y}_{a.c}}$ and ${\hat{y}_{h.c}}$ are outputs of the sub-network for the arm-reaching MI and hand-related MI classification. The number of arm-reaching classes determines the parameter \textit{M}.

\begin{figure}[!t]
  \centerline{\includegraphics[width = \columnwidth]{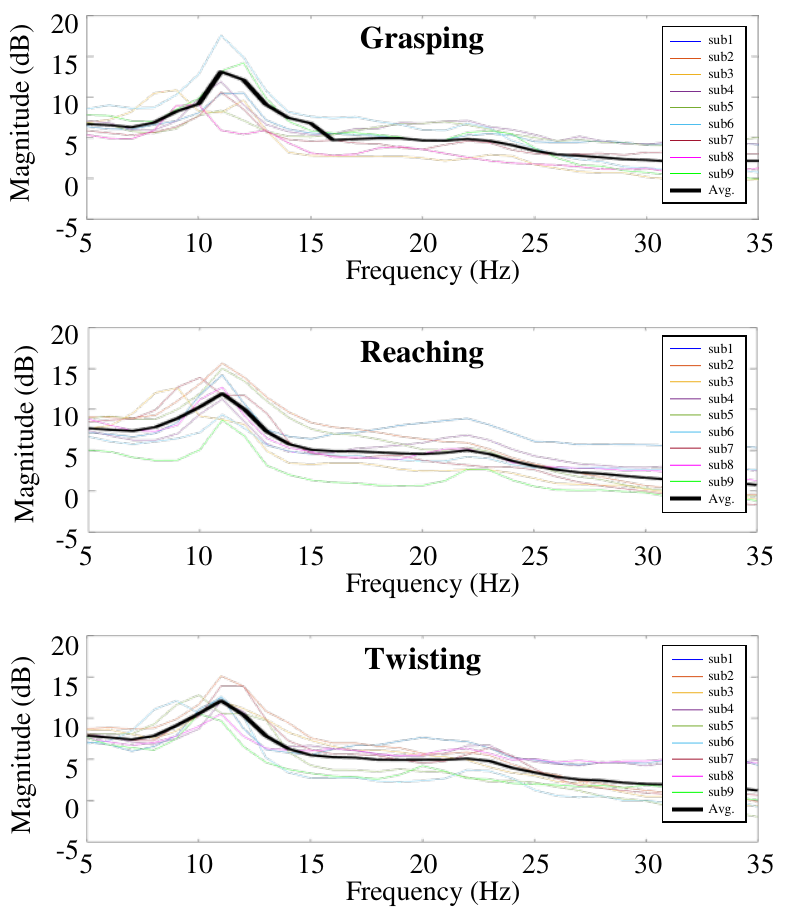}}
  \caption{Power spectral density per each subject and average of across all subjects (black line)}
\end{figure}
\section{Results and Discussion}
\label{sec:print}

Fig. 3 shows the power spectral density (PSD) of EEG data per MI task. The frequency-domain analysis was conducted on EEG data for seeking frequency characteristics using PSD. Through the analysis, we confirmed that a high magnitude was obtained in mu-band (8-12 Hz). Accordingly, it is advantageous to extract frequency features or band power features of the EEG data. For evaluation, the data was organized into 3-class, 5-class, and 7-class. The dataset for 3-class classification contains categorized arm-reaching MI, hand-related MI, and resting state classes. For 5-class classification, we comprised the dataset with arm-reaching (left and right), grasping, twisting, and resting state. The dataset for 7-class classification further separated vertical reaching (VER) classification and horizontal reaching (HOR) classification. We added upward, downward, forward, and backward to the arm-reaching classes in the 5-class dataset. We used 32-size mini-batch and 200 epochs for training. For comparison with existing MI classification approaches, all experiments were conducted in the same test environment.

Table 2 indicates the classification accuracies of the ERA-CNN for each dataset. The dataset was split into training and test data for evaluation. In the 3-class classification, we used only a shared layer since the dataset consists of three categorized classes. As shown in Table 2, both 7-class dataset classification performances are slightly different (0.63 and 0.66). The highest accuracy of VER classification is 0.80 in sub8. On the other hand, the highest accuracy of HOR classification is 0.68 in sub1 and sub8. The chance level of the 7-class classification is around 0.14. We found that almost subjects who performed well in 3-class tended to show higher classification performance in other classifications.
                         
Table 3 shows a comparison of classification accuracies and standard deviations with existing methods. EPA-CNN outperformed comparison groups in the classification accuracy. Unlike other methods based on singular structure, ERA-CNN divides the classes according to their roles. Hence, it classifies a series of a small number of classification classes, which can explain the increase in performance over singular models that classify entire classes at once. However, the ShallowConvNet which has a similar architecture with the shared layer marks the second-highest accuracy (0.78) in 3-class classification. Even in 7-class classifications, the accuracy of the ShallowConvNet shows better performance (0.42 and 0.48) than other methods because it extracts frequency band power features like the ERA-CNN. However, in the 5-class classification, all three methods record similar accuracies except FBCSP with RLDA. The difference in overall classification accuracy of the ERA-CNN model over existing methods was found to be significant using a paired \textit{t}-test (\textit{p}-value $<$ 0.05). Due to the relatively small size of the dataset compared to other domains, performance can be improved using either augmentation or sliding window methods.
%

\begin{table}[t!]
\caption{Classification results per each classification}
\renewcommand{\arraystretch}{1.1}
\resizebox{\columnwidth}{!}{%
\begin{tabular}{ccccc} \hline
Subjects      &  3-class & 5-class & 7-class (HOR) & 7-class (VER) \\ \hline
sub1  & 0.93    & 0.84    & \textbf{0.68}          & 0.64         \\
sub2  & \textbf{0.96}    & \textbf{0.86}    & 0.67         & 0.68        \\
sub3  & 0.82    & 0.78    & 0.65         & 0.74         \\
sub4  & 0.88    & \textbf{0.86}    & 0.65         & 0.64         \\
sub5  & 0.78    & 0.76    & 0.56         & 0.54         \\
sub6  & 0.90    & 0.78    & 0.62         & 0.71         \\
sub7  & 0.88   & 0.82    & 0.57         & 0.58         \\
sub8  & 0.90    & \textbf{0.86}    & \textbf{0.68}         & \textbf{0.80} \\
sub9  & 0.88    & 0.84    & 0.58        & 0.62         \\ \hline
Avg. & 0.88   & 0.82   & 0.63         & 0.66         \\
Std.  & 0.05    & 0.03    & 0.04          & 0.07      \\   \hline
\end{tabular}}%
\end{table}


\begin{table}[t!]
\caption{Comparison with existing methods} 
\renewcommand{\arraystretch}{1.23}
\resizebox{\columnwidth}{!}{%
\begin{tabular}{lcccc} \hline
Methods               & 3-class       & 5-class       & 7-class (HOR) & 7-class (VER) \\ \hline
\textit{FBCSP+RLDA} \cite{channel}   & 0.44 (0.08)  & 0.44 (0.05)  & 0.27 (0.04)  & 0.26 (0.04)  \\
\textit{DeepConvNet} \cite{deepconvnet}   & 0.70 (0.09)   & 0.56 (0.09)  & 0.37 (0.09)  & 0.39 (0.07)  \\
\textit{ShallowConvNet} \cite{deepconvnet} & 0.78 (0.10) & 0.54 (0.11) & 0.42 (0.08)  & 0.48 (0.08)  \\
\textit{EEGNet} \cite{EEGNET}       & 0.69 (0.13) & 0.55 (0.08)  & 0.36 (0.06)  & 0.41 (0.08)  \\
\textit{\textbf{ERA-CNN}}        & \textbf{0.88 (0.05)}  & \textbf{0.82 (0.03)}  &\textbf{0.63 (0.04)}  &\textbf{0.66 (0.07)}  \\   \hline
\end{tabular}%
}
\end{table}

\section{Conclusion and Future works}
\label{sec:print}
In this paper, we proposed an ERA-CNN architecture that considers discriminative features for each upper limb region of MI classification. We demonstrated that the ERA-CNN achieved the highest classification accuracies (0.86, 0.82, 0.63, and 0.66) compared to existing methods (0.76, 0.56, 0.42, and 0.48). This improvement in performance opens up the possibility to perform continuous decoding for various types of upper limb movements. The proposed model can thus be applied to help intuitively control external devices with high accuracy, such as a robotic arm, which can ultimately help improve the autonomy of people with movement disabilities.


\section{Acknowledgement}
The authors thanks to B.-H. Kwon and J.-H. Cho for their help with the dataset construction and S. K. Prabhakar, P. Bertens and J. Kalafatovich for their discussion of the data analysis.





\bibliographystyle{IEEEbib}
\bibliography{strings,refs}

\end{document}